\documentclass[oribibl]{llncs}

\PassOptionsToPackage{hyphens}{url}
\usepackage[colorlinks = true,
  linkcolor = blue,
  citecolor = blue,
  urlcolor = blue]{hyperref}

\usepackage{url}
\usepackage{csquotes}
\usepackage{multirow}
\usepackage{amsmath}
\usepackage{amssymb}
\usepackage{graphicx}
\usepackage{tabularx}
\usepackage{booktabs}
\usepackage{threeparttable}

\begin{document}

\title{Vulnerability Patching Across Software \\ Products and Software Components:
  A Case Study of Red Hat's Product Portfolio}

\author{Jukka Ruohonen\orcidID{0000-0001-5147-3084} \and \\ Sani Abdullahi\orcidID{0000-0003-4962-2794} \and \\ Abhishek Tiwari\orcidID{0000-0001-8415-5410} \\ \email{\{juk, saa, abti\}@mmmi.sdu.dk}}
\institute{University of Southern Denmark, S\o{}nderborg \& Odense, Denmark}

\maketitle

\begin{abstract}
Motivated by software maintenance and the more recent concept of security debt,
the paper presents a time series analysis of vulnerability patching of Red Hat's
products and components between 1999 and 2024. According to the results based on
segmented regression analysis, the amounts of vulnerable products and components
have not been stable; a linear trend describes many of the series well. Nor do
the amounts align well with trends characterizing vulnerabilities in
general. There are also visible breakpoints indicating that the linear trend is
not universally applicable and that the growing security debt may be
stabilizing.
\end{abstract}


\section{Introduction}

Software maintenance is a classical topic in software engineering and software
security. Vulnerabilities must be patched, among other things. Yet, at the same
time, software maintenance is costly due to the human effort required. Although
vulnerabilities and other security issues are often prioritized,
\text{developers---and} perhaps particularly open source software
developers---have sighed about a general maintenance burden~\cite{Linaker24}. By
hypothesis, such a burden is related to a software's characteristics, including
its size and complexity, which are typically (but perhaps not necessarily)
strongly correlated~\cite{Mamun17}. As the paper continues existing empirical
studies~\cite{Alfasi24, Alhazmi05, Ruohonen15COSE, Xu22} by operating with a
large software portfolio of Red Hat, which has been a subsidiary of IBM since
2019, the points about software size and software complexity apply well as a
general motivation. In other words, the paper's context is about a large-scale
and long-lived open source software stack maintained by a commercial company
with respect to security patches delivered particularly to its paid
customers. This motivating point can be further strengthened by recalling that
the Red Hat's flagship product is an open source software operating system with
a support period of ten years or even more.

To sharpen the theoretical motivation a little, it can be also said that the
paper presents a time series analysis of security maintenance involving Red
Hat's product portfolio from a perspective of software \textit{security
  debt}. Although not a well-defined concept, software security debt has been
seen to be a subset of technical debt, and software vulnerabilities have been
perceived as a typical (but not the only) indicator of software security
debt~\cite{Ferreyra24, Rindell19a, Rindell19b, Sion24}.\footnote{Based on an
industry case study, a definition was recently proposed: security ``debt is a
set of design or implementation solutions that hinder or has the potential to
hinder the achievement of a system's security goal''
\cite[p.~50]{Kruke24}. Related literature reviews~\cite{Huopio20, Martinez21}
summarize different characteristics of security debt and metrics for~it.}  If
vulnerabilities are continuously disclosed, coordinated, investigated, and
patched, a software could be argued to have a security debt problem. Like all
debts, it should be ideally paid back. However, the case of Red Hat changes the
interpretation a~little.

Although Red Hat develops a lot of software also by itself, practically all of
it is open source software and most of the company's products build upon
numerous open source software components. This characterization means that a
potential security debt may well not be of Red Hat's making but instead
originating from the open source software projects whose software is used and
distributed by Red Hat. Although the dataset soon described does not allow to
delve deeper, a sensible corollary hypothesis for further research would be that
a growing size of Red Hat's open source software portfolio would imply also a
growing software security debt. Given the explosion of disclosed and archived
vulnerabilities, the growth rate of which seems to follow an exponential
trend~\cite{Wochnik25}, also other factors are at play, including a potentially
increased pool of people looking for vulnerabilities as well as potentially
improved detection tools.

With these introductory notes, the first research question (RQ.1) examined is:
\textit{how many Red Hat products and their software components have been
  impacted by vulnerabilities on average over the years?} Answer to this
research question already sheds some light on the overall hypothetical
longitudinal security debt present in the Red Hat's product portfolio. The
second research question (RQ.2) brightens the light: \textit{have the amounts
  remained longitudinally stable?} Although this RQ.2 only speaks about
longitudinal stability, the implicit hypothesis is that there have been growth
trends proxying increased security debt over the years. The third and final
research question (RQ.3) asks \textit{whether the amounts and their longitudinal
  stability or instability correspond with the stability or instability of the
  total number of vulnerabilities processed?} Although only tentatively, an
answer to RQ.3 sheds light on whether the vulnerabilities specific to the Red
Hat's product portfolio follow the general trend of vulnerabilities reported.

As for the paper's remaining structure, the data and methods are presented in
Sections~\ref{sec: data} and \ref{sec: methods}, respectively. The former
section also further clarifies how the concepts of products and components are
operationalized, whereas the latter section elaborates how some of the
qualifying terms in the RQs are interpreted. Then, results are presented in
Section~\ref{sec: results}. The final Section~\ref{sec: conclusion} concludes.

\section{Data}\label{sec: data}

The empirical dataset covers archived vulnerabilities for the Red Hat's product
portfolio from the first day of January 1999 to the last day of December
2024. This period ensures that full annual records are present for each
year. Among the well-known products in the portfolio is the Red Hat Enterprise
Linux (RHEL) that was initially released already twenty-five years ago. Since
then, Red Hat has extended its product portfolio toward many other areas as
well, including, but not limited to, software development tools and frameworks,
artificial intelligence, and cloud computing applications~\cite{RedHat25a}. In
line with the RQs, however, the focus is not on any particular products
\textit{per~se} but rather on the amounts of products and their components that
were supported and maintained by Red Hat at a given date. These are meticulously
cataloged by Red Hat regarding vulnerabilities. For further improving the
cataloging and tracking, the company migrated its vulnerability data into a new
Vulnerability Exploitability eXchange (VEX) format in the early
2020s~\cite{RedHat23}. The dataset was assembled from a VEX-based archival file
downloaded on 17 August 2025 from a Red Hat's repository~\cite{RedHat25b}. The
dataset used is also available online for replication
purposes~\cite{Ruohonen25DSVEX}. All vulnerabilities tracked by Red Hat are
identified by Common Vulnerabilities and Exposures (CVEs). The VEX format
specifies four statuses for the vulnerabilities.

The ``\textit{fixed}'' status denotes that a particular vulnerability was
patched in a given product and its given components. The ``\textit{known
  affected}'' status is similar but it references cases in which products and
components are confirmed to be affected by a particular vulnerability but no fix
is yet available. Both statuses provides rather straightforward metrics to
investigate the hypothesized security debt. The ``\textit{under investigation}''
status is about security debt indirectly; it refers to CVEs whose applicability
to the Red Hat's product portfolio was under an investigation at a given
date. Since an investigation, including a verification, also represent
maintenance work, this status too can be interpreted to represent security
debt. If a given CVE was not applicable, work was still required. Given that
postponing vulnerability patching has been seen to increase security
debt~\cite{Kruke24}, a~growth trend in a metric representing the ``\textit{under
  investigation}'' status would also have a relatively straightforward
interpretation.

It should be further emphasized that the three statuses are potentially related
to each other. If a vulnerability was under an investigation at a given date, it
may have ended having the ``\textit{known affected}'' status later on in case
the vulnerability was confirmed but a patch for it was difficult to backport or
develop in-house. At a later date it may have also attained a ``\textit{fixed}''
status or remain unfixed. These considerations are also related to the embargo
periods Red Hat enforces for limiting unnecessary visibility during
vulnerability patching~\cite{Lin23}. Although the VEX format provides crude
bookkeeping material on the changes made, the potential relations of the three
statutes imply that the results reported only apply to the data collection date
noted earlier. Finally, the ``\textit{known not affected}'' status, which
provides a confirmation that some specific products and components were not
affected by a given CVE, is not suitable for a time series analysis because many
of the entries with this status merely provide a character string
\texttt{red\_had\_products}. In other words, the counts specified for RQ.1 and
RQ.2 remain unknown because it cannot be deduced how many products and their
components Red Hat had and maintained at a given date.

The three statuses used---henceforth, fixed, affected, and investigated---were
parsed from the \texttt{product\_status} field under the
\texttt{vulnerabilities} field in the online archival file delivered in the
JavaScript Object Notation (JSON) format. Also the paper's terminology follows
the JSON-represented VEX format, which explicitly talks about products and
components, further emphasizing that the latter contains also other components
than traditional Linux packages~\cite{RedHat24a}. In practice, parsing was done
by using a colon to split packages from components; some entries have also a
third colon-separated field representing instruction set architectures but it
was ignored without a particular loss of relevant details. Finally, the monthly
and yearly aggregates refer to the numbers of unique products and components
that were affected by unique CVEs at a particular month or year. Regarding RQ.3,
the monthly and yearly aggregates refer to the unique CVEs present in the
archival file used, regardless of their particular statuses.

\section{Methods}\label{sec: methods}

The analysis is carried out separately by using monthly and yearly
aggregates. Although robust statistics, such as median, are suitable for
initially answering to RQ.1, strong growth trends may also make robust
statistics misleading longitudinally. As the interest is not to examine which
particular trend might best characterize the time series, a linear trend is as a
good trend as any for a starting point. To this end, a bootstrap-based
estimation for testing a presence of a linear trend is used. Specifically, the
null hypothesis is that $\hat{\beta} \approx 0$ in a $y_t = \alpha + \beta t +
\epsilon_t$, where $y_t$ is a given series, $t = 1, \ldots, T$ denotes either
days, months, or years and $T$ the length of a given time series, $\alpha$ is a
constant, $\beta$ is a regression coefficient, and $\epsilon_t$ is a residual
term following a $k$-order autoregressive (AR) process (for further details see
\cite{Noguchi11} and the implementation \cite{funtimes} used for
computation). If the null hypothesis is rejected, the answer to RQ.2 is negative
and the ``on average'' wording used for RQ.1 attains a slightly different
interpretation.

To examine the wording ``stable'' used for RQ.2, segmented ordinary least
squares (OLS) and Poisson regressions are estimated using an existing
implementation~\cite{Muggeo08}. As previously, the only independent metric is
the linear trend. And as worded in the implementation, the interest is to
examine ``broken linear trends'' around breakpoints estimated from data~(for
details see \cite{Muggeo16}). The general time series background of the
segmented regression is closely related to the concepts of structural change,
breakpoints, and parameter stability~(cf.~\cite{Ruohonen25ICTSSa}). For
simplicity, only a single breakpoint is specified. Therefore, a visually
forceful breakpoint generally also means that a linear trend could be seen to
contain two distinct slope coefficients, $\hat{\beta}_1$ and $\hat{\beta_2}$. If
these are large in magnitude and a coefficient of determination~(R$^2$) is large
too, also the formal test described earlier should reject its null
hypothesis. If $\hat{\beta}_1$ and $\hat{\beta}_2$ are further very different
from each other in terms of their magnitudes, a negative answer to RQ.2 attains
further weight; security debt has been present. That said---and as will be seen,
already a visual interpretation is sufficient for answering to RQ.2 and RQ.3.

\section{Results}\label{sec: results}

The dissemination of the time series results can be started with the formal
linear trend testing. So, the test noted earlier rejects the null hypothesis of
no linear trend at a \text{$p < 0.05$} level for all series. Changing the
automatic determination of the AR($k$) orders to manual picks does not change
the conclusion, as can be also seen from the $k = 2$ cases in Table~\ref{tab:
  trend test}. Also the monthly and yearly CVE series favor the alternative
hypothesis at the same statistical significance level. All series can be
concluded to be trending, although the test results say nothing about whether a
linear trend is particularly suitable.

\begin{table*}[th!b]
\centering
\caption{Linear Trend Tests with AR(2) Terms}
\label{tab: trend test}
\begin{tabular}{llrrcrr}
\toprule
Aggregation\qquad\qquad & Series\qquad\qquad\qquad
& \multicolumn{2}{c}{Products}
&& \multicolumn{2}{c}{Components} \\
\cmidrule{3-4}\cmidrule{6-7}
&& $t$-value &\qquad $p$-value &\qquad\qquad& $t$-value &\qquad $p$-value \\
\hline
Yearly & Fixed & $13.856$ & $0.009$ && $18.625$ & $< 0.001$ \\
Monthly & Fixed & $28.975$ & $< 0.001$ && $23.774$ & $< 0.001$ \\
\cmidrule{3-7}
Yearly & Affected & $20.116$ & $< 0.001$ && $8.675$ & $< 0.010$ \\
Monthly & Affected & $36.178$ & $< 0.001$ && $18.352$ & $< 0.001$ \\
\cmidrule{3-7}
Yearly & Investigated & $4.454$ & $0.009$ && $4.287$ & $0.017$ \\
Monthly & Investigated & $7.264$ & $< 0.001$ && $6.119$ & $< 0.001$ \\
\bottomrule
\end{tabular}
\end{table*}

\begin{table*}[th!b]
\centering
\caption{Performance of Segmented OLS Regressions}
\label{tab: performance}
\begin{tabular}{llrrcrr}
\toprule
Aggregation\qquad\qquad & Series\qquad\qquad\qquad
& \multicolumn{2}{c}{Products}
&& \multicolumn{2}{c}{Components} \\
\cmidrule{3-4}\cmidrule{6-7}
&& R$^2$ &\qquad BIC &\qquad\qquad& R$^2$ &\qquad BIC \\
\hline
Yearly & Fixed &  $0.927$ & $304$ && $0.958$ & $412$ \\
Monthly & Fixed & $0.754$ & $3238$ && $0.664$ & $4630$ \\
\cmidrule{3-7}
Yearly & Affected & $0.973$ & $201$ && $0.973$ & $357$ \\
Monthly & Affected & $0.879$ & $2306$ && $0.660$ & $3984$ \\
\cmidrule{3-7}
Yearly & Investigated & $0.757$ & $189$ && $0.880$ & $252$ \\
Monthly & Investigated & $0.192$ & $1588$ && $0.180$ & $2496$ \\
\bottomrule
\end{tabular}
\end{table*}

\begin{figure}[t!]
\centering
\includegraphics[width=\linewidth, height=4cm]{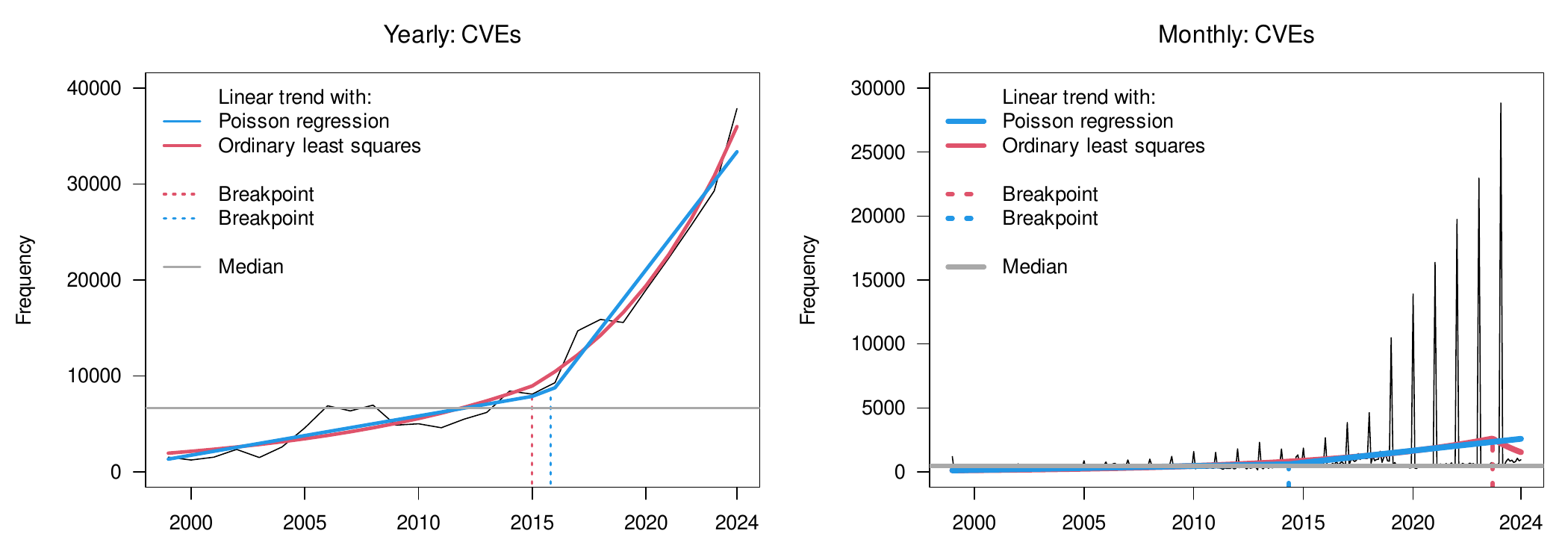}
\caption{CVEs Processed by Red Hat}
\label{fig: cves}
\end{figure}

Therefore, Table~\ref{tab: performance} shows a summary of the performance of
the segmented OLS regressions estimated with a linear trend. In addition to the
coefficients of determination, Bayesian information criterion (BIC) values are
shown; lower values are better. As can be seen, the linear trend with single
breakpoints do provide rather good estimates particularly for the yearly
aggregates. The yearly aggregates for the affected status series attain the
highest $R^2$-value of $0.973$ for both products and components. The monthly
aggregates show lesser performance, which is expected as the frequency and
fluctuations, including potential periodicity, are higher than in the yearly
aggregates. Even then, the monthly aggregates for the fixed and affected
statuses are still moderately decent. The monthly aggregates for the
investigated status series are outliers, indicating that a linear trend is not
suitable for them.

\begin{figure}[t!]
\centering
\includegraphics[width=\linewidth, height=8cm]{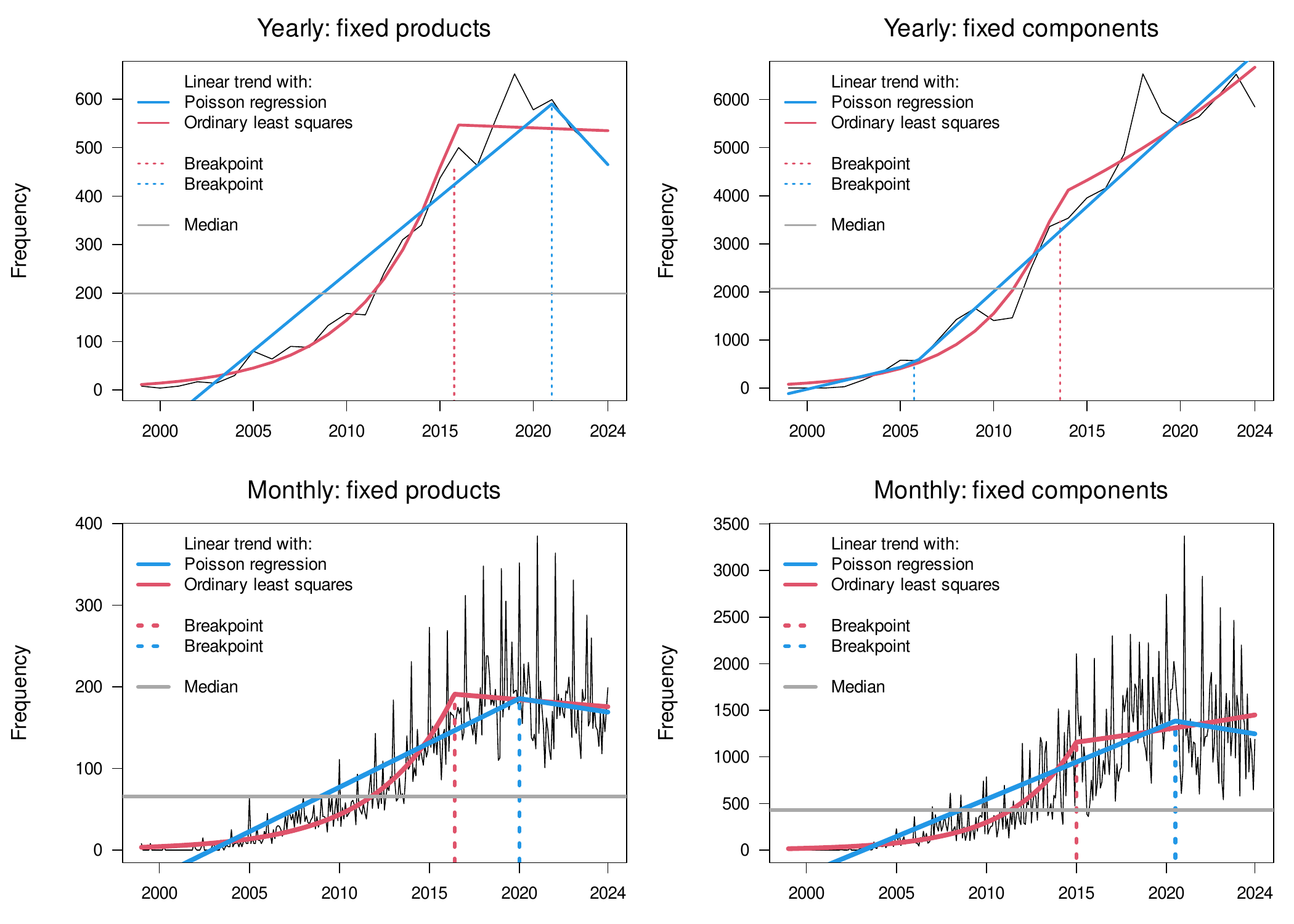}
\caption{Fixed Products and Components}
\label{fig: fixed}
\end{figure}
\begin{figure}[t!]
\centering
\includegraphics[width=\linewidth, height=8cm]{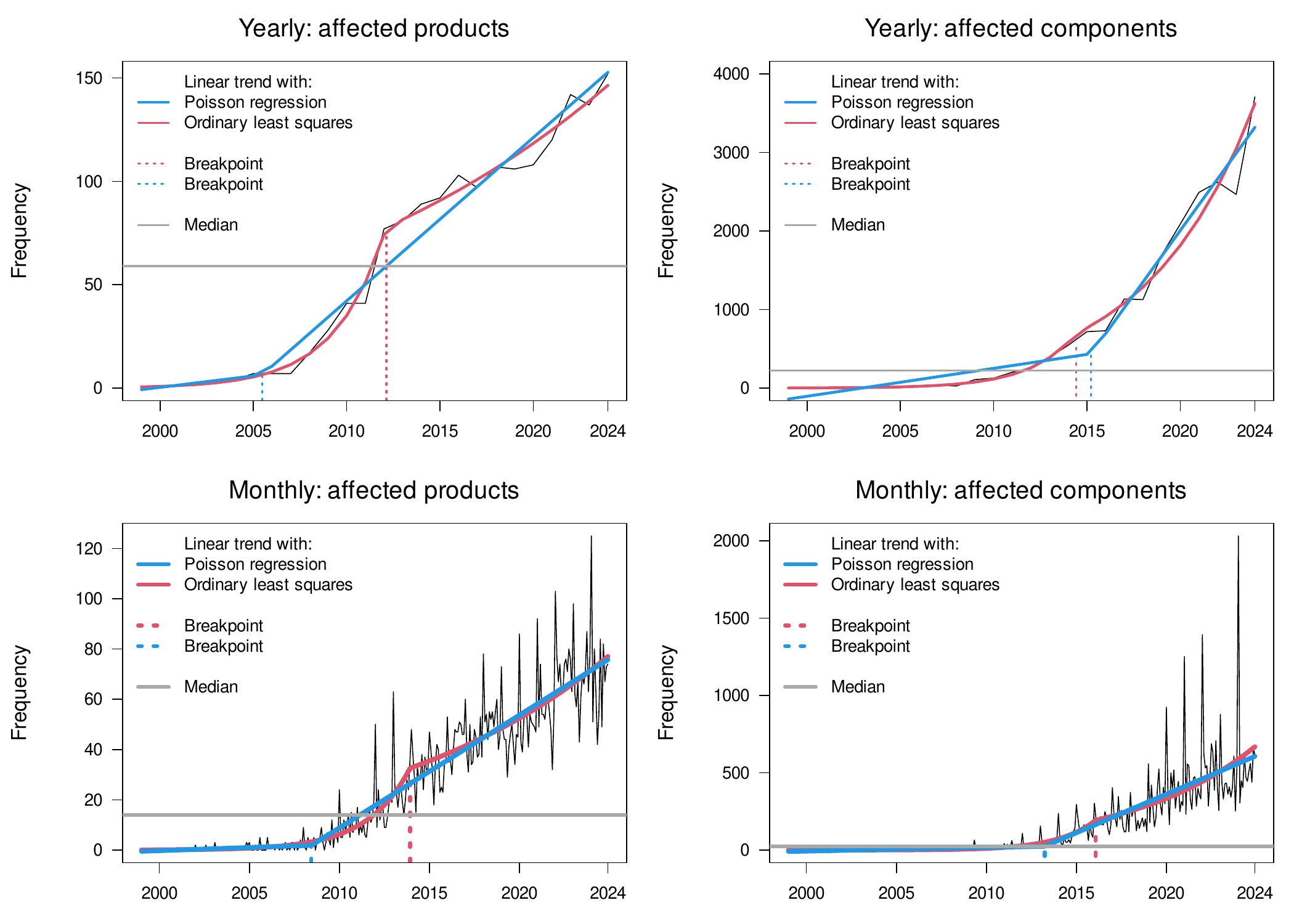}
\caption{Affected Products and Components}
\label{fig: affected}
\end{figure}
\begin{figure}[t!]
\centering
\includegraphics[width=\linewidth, height=8cm]{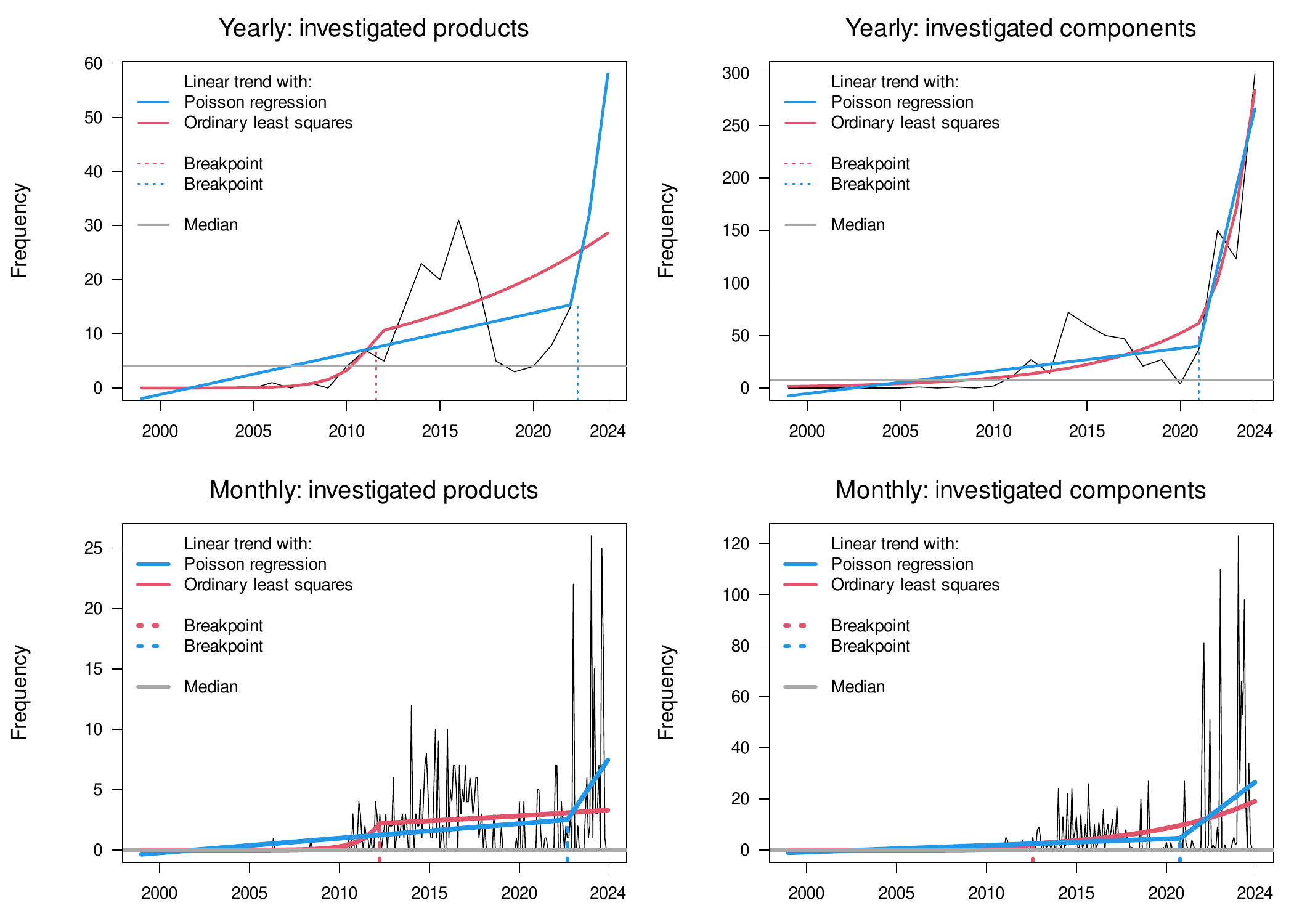}
\caption{Investigated Products and Components}
\label{fig: investigated}
\end{figure}

Before continuing to analyze the segmented regression results further, a look at
the CVE series can be done beforehand for preparing an answer to RQ.3. Thus,
Fig.~\ref{fig: cves} shows and visualizes the results for the CVE series. As can
be seen from the left-hand side plot, a linear trend provides a very good
estimate for the yearly aggregates; the segmented OLS regression yields a
$\textmd{R}^2 \approx 0.969$. Also the Poisson regression provides a good
estimate. It can be also seen that the breakpoints detected are around 2015
after which the amount of CVEs processed by Red Hat has accelerated. Given that
earlier research operating with a period between 1990 and 2015 has seen the year
$2005$ as an inflection point in the acceleration of CVEs publicly
archived~\cite{Ruohonen16GIQ}, it seems that the inflection point has moved
alongside the even more rapid acceleration after around 2015. In any case, it is
also important to stress that a linear trend is ill-suited for describing the
monthly CVE series; for this series, $\textmd{R}^2$ is as low as approximately
$0.072$. When taking a look at the right-hand side plot, the reason is clear; a
strong periodicity is present in the monthly CVE series. A plausible explanation
for the periodicity is related to release engineering schedules. In this regard,
existing results regarding periodicity~\cite{Joh17} align rather closely with
the Microsoft's famous ``patch Tuesday'' policy~\cite{Morrissey23}. Although Red
Hat does not have a similar policy regarding patches, minor releases for RHEL
are done every six months~\cite{RedHat24b}. Further research would be needed for
a definite answer---especially since related time series results are not
entirely confirmatory~\cite{Ruohonen15IWPSE}, but for the time being it suffices
to tentatively say that the RHEL's release engineering schedule may well be
behind the periodicity visible in the right-hand side plot of Fig.~\ref{fig:
  cves}.

However, none of the other series exhibit notable periodicity. This observation
can be seen from Figs.~\ref{fig: fixed}, \ref{fig: affected}, and \ref{fig:
  investigated}, which visualize the results for the remaining series. When
looking at the plots in the three figures, the $y$-axes should be kept in
mind. In other words, most of the CVEs processed by Red Hat are about fixed
vulnerabilities; the amounts with the affected status are much lower, and the
amounts with the investigated status even lower. In other words, most of the
vulnerabilities have been patched. Although $58$ products and $299$ components
were still being investigated in 2024, the amounts are still small when compared
to the $466$ products and $5,847$ components that were already patched in
2024. With this point in mind, the visualizations clearly reiterate the negative
answer to RQ.2. All series are trending, some more and some less. Regarding
RQ.1, the upper-left plot in Fig.~\ref{fig: fixed} seems suitable for
summarizing an answer because it indicates a stabilization or even a flattening
of the linear growth curve from about 2016 onward. The medians between 2016 and
2024 are $542$ products and $5,731$ components with a fixed status. Although the
upper-right plot indicates no stabilization or flattening in terms of vulnerable
components fixed, on one hand, it could be said in a tentative manner that the
security debt has slightly decreased in the early 2020s. On the other hand, when
keeping the points made in Section~\ref{sec: data} in mind, some or even most of
the amounts in Figs.~\ref{fig: affected} and \ref{fig: investigated} will likely
end up having a fixed status, meaning that the growth likely continues also in
the nearby future. Finally, regarding RQ.3, the slopes of the linear trends
indicate that the growth has started earlier than with the CVE series in
Fig.~\ref{fig: cves}. The yearly aggregates for the fixed products indicate an
inflection point already before 2015. In other words, the answer to RQ.3 is
negative and the results specific to Red Hat do not align well with previous
studies on CVEs in general.

\section{Conclusion}\label{sec: conclusion}

The paper presented a time series analysis of vulnerability patching involving
Red Hat's software products and components. Three research questions were
specified to guide the analysis. \textit{If fixed products and components are
  considered, between 2016 and 2024 over five hundred products and nearly six
  thousand components were patched on average} (RQ.1). However, \textit{the
  amounts of products and components patched have not been stable over the
  years} (RQ.2). All time series investigated are trending, some more and some
less. A linear trend with breakpoints yield good statistical performance in
overall. \textit{The trends and their breakpoints also indicate that the trends
  characterizing patching of the Red Hat's product portfolio are not equal to
  those characterizing CVEs in general}~(RQ.3). Although a verification check is
needed in the future, the breakpoints also hint that the growing security debt
might be slowly stabilizing. If an explicit goal would be to minimize it,
perhaps an obvious starting point might be to reduce the size of the software
portfolio. This point is also a limitation of the paper.

In other words, the analysis did not consider how large and complex the Red
Hat's portfolio actually is. As discussed in existing research~\cite{Wochnik25},
the point is also theoretically relevant because the patching trends observed
should be ideally compared to other trends, including those characterizing the
supposedly growing size of the portfolio. Regarding limitations more generally,
the many problems with vulnerability databases should be acknowledged as a
potential threat to validity too. Red Hat's vulnerability data is not an
exception~\cite{Alfasi24, Xu22}. Addressing these limitations would offer a good
topic for further research. In terms of research implications, the generally
good performance of the segmented regressions is worth emphasizing. As closely
related research has often operated with non-linear models~\cite{Alhazmi05,
  Ruohonen15COSE}, the results can be used to argue that simpler models may work
equally well. After all, simplicity brings easier~interpretation.

\bibliographystyle{splncs03}

\begin{thebibliography}{10}
\providecommand{\url}[1]{\texttt{#1}}
\providecommand{\urlprefix}{URL }

\bibitem{Mamun17}
{Al Mamun}, M.A., Berger, C., Hansson, J.: {C}orrelations of {S}oftware {C}ode
  {M}etrics: {A}n {E}mpirical {S}tudy. In: Proceedings of the 27th
  International Workshop on Software Measurement and 12th International
  Conference on Software Process and Product Measurement (IWSM Mensura 2017).
  pp. 255--266. ACM, Gothenburg (2017)

\bibitem{Alfasi24}
Alfasi, D., Shapira, T., Bremler-Barr, A.: {V}uln{S}copper: {U}nveiling
  {H}idden {L}inks {B}etween {U}nseen {S}ecurity {E}ntities. In: Proceedings of
  the 3rd GNNet Workshop on Graph Neural Networking Workshop (GNNet 2024). pp.
  33--40. ACM, Los Angeles (2024)

\bibitem{Alhazmi05}
Alhazmi, O., Malaiya, Y., Ray, I.: {S}ecurity {V}ulnerabilities in {S}oftware
  {S}ystems: {A} {Q}uantitative {P}erspective. In: Proceedings of the 19th
  Annual IFIP WG 11.3 Working Conference on Data and Applications Security
  (DBSec 2005). pp. 281--294. Springer (2005)

\bibitem{Ferreyra24}
Ferreyra, N.E.D., Shahin, M., Zahedi, M., Quadri, S., Scandariato, R.: {W}hat
  {C}an {S}elf-{A}dmitted {T}echnical {D}ebt {T}ell {U}s {A}bout {S}ecurity?
  {A} {M}ixed-{M}ethods {S}tudy. In: Proceedings of the 21st International
  Conference on Mining Software Repositories (MSR 2024). pp. 704--715. ACM,
  Lisbon (2024)

\bibitem{Huopio20}
Huopio, S.: {A} {Q}uest for {I}ndicators of {S}ecurity {D}ebt. The Cyber
  Defense Review  5(1),  169--184 (2020)

\bibitem{Joh17}
Joh, H., Malaiya, Y.K.: {P}eriodicity in {S}oftware {V}ulnerability
  {D}iscovery, {P}atching and {E}xploitation. International Journal of
  Information Security  16,  673--690 (2017)

\bibitem{Kruke24}
Kruke, M.M., Martini, A., Cruzes, D.S., Iovan, M.: {D}efining {S}ecurity
  {D}ebt: {A} {C}ase {S}tudy {B}ased on {P}ractice. In: Proceedings of the 25th
  International Conference on Product-Focused Software Process Improvement
  (PROFES 2024). pp. 43--59. Sprigner, Tartu (2024)

\bibitem{Lin23}
Lin, J., Zhang, H., Adams, B., Hassan, A.E.: {V}ulnerability {M}anagement in
  {L}inux {D}istributions: {A}n {E}mpirical {S}tudy on {D}ebian and {F}edora.
  Empirical Software Enigineering  28,  1--34 (2023)

\bibitem{Linaker24}
Lin\r{a}ker, J., Link, G.J.P., Lumbard, K.: {S}ustaining {M}aintenance {L}abor
  for {H}ealthy {O}pen {S}ource {S}oftware {P}rojects {T}hrough {H}uman
  {I}nfrastructure: {A} {M}aintainer {P}erspective. In: Proceedings of the 18th
  ACM/IEEE International Symposium on Empirical Software Engineering and
  Measurement (ESEM 2024). pp. 37--48. ACM, Barcelona (2024)

\bibitem{funtimes}
Lyubchich, V., Gel, Y.R., Brenning, A., Chu, C., Huang, X., Islambekov, U.,
  Niamkova, P., Ofori-Boateng, D., Schaeffer, E.D., Vishwakarma, S., Wang, X.:
  funtimes: {F}unctions for {T}ime {S}eries {A}nalysis (2023), {R} package
  version 9.1, available online in August 2025:
  \url{https://cran.r-project.org/web/packages/funtimes/index.html}

\bibitem{Martinez21}
Martinez, J., Quintano, N., Ruiz, A., Santamaria, I., {de Soria}, I.M., Arias,
  J.: {S}ecurity {D}ebt: {C}haracteristics, {P}roduct {L}ife-{C}ycle
  {I}ntegration and {I}tems. In: Proceedings of the IEEE/ACM International
  Conference on Technical Debt (TechDebt 2021). pp. 1--5. IEEE, Madrid (2021)

\bibitem{Morrissey23}
Morrissey, C.: {W}indows {M}onthly {U}pdates {E}xplained (2023), {M}icrosoft
  {W}indows {IT} {P}ro {B}log, available online in August 2025:
  \url{https://techcommunity.microsoft.com/blog/windows-itpro-blog/windows-monthly-updates-explained/3773544}

\bibitem{Muggeo08}
Muggeo, V.M.R.: segmented: {A}n {R} {P}ackage to {F}it {R}egression {M}odels
  with {B}roken-{L}ine {R}elationships. R News  8,  20--25 (2008)

\bibitem{Muggeo16}
Muggeo, V.M.R.: {T}esting {W}ith a {N}uisance {P}arameter {P}resent {O}nly
  {U}nder the {A}lternative: {A} {S}core-{B}ased {A}pproach {W}ith
  {A}pplication to {S}egmented {M}odelling. Journal of Statistical Computation
  and Simulation  86(15),  3059--3067 (2016)

\bibitem{Noguchi11}
Noguchi, K., Gel, Y.R., Duguay, C.R.: {B}ootstrap-{B}ased {T}ests for {T}rends
  in {H}ydrological {T}ime {S}eries, {W}ith {A}pplication to {I}ce {P}henology
  {D}ata. Journal of Hydrology  410,  150--161 (2011)

\bibitem{RedHat24a}
{Red Hat}: {CSAF} {S}ecurity {A}dvisories and {VEX} {S}ecurity {D}ata (2024),
  {R}ed {H}at {S}ecurity {D}ata {G}uidelines, available online in August 2025:
  \url{https://redhatproductsecurity.github.io/security-data-guidelines/csaf-vex/#csaf-security-advisories-and-vex-security-data}

\bibitem{RedHat24b}
{Red Hat}: {H}ow {O}ften {R}ed {H}at {R}eleases the {RHEL} {P}atches and
  {U}pgrades? (2024), {R}ed {H}at {C}ustomer {P}ortal, available online in
  August 2025: \url{https://access.redhat.com/solutions/3711551}

\bibitem{RedHat25a}
{Red Hat}: {A}ll {R}ed {H}at {P}roducts (2025), {A}vailable online in August
  2025: \url{https://www.redhat.com/en/technologies/all-products}

\bibitem{RedHat25b}
{Red Hat}: {I}ndex of /csaf/v2/vex/ (2025), csaf\_vex\_2025-08-16.tar.zst,
  7e4f71aa31bbb75188c4d337f1eadee7ed76b213 (SHA1), available online in August
  2025: \url{https://security.access.redhat.com/data/csaf/v2/vex/}

\bibitem{Rindell19a}
Rindell, K., Bernsmed, K., Jaatun, M.G.: {M}anaging {S}ecurity in {S}oftware:
  {O}r: {H}ow {I} {L}earned to {S}top {W}orrying and {M}anage the {S}ecurity
  {T}echnical {D}ebt. In: Proceedings of the 14th International Conference on
  Availability, Reliability and Security (ARES 2019). pp. 1--8. ACM, Canterbury
  (2019)

\bibitem{Rindell19b}
Rindell, K., Holvitie, J.: {S}ecurity {R}isk {A}ssessment and {M}anagement as
  {T}echnical {D}ebt. In: Proceedings of the International Conference on Cyber
  Security and Protection of Digital Services (Cyber Security 2019). pp. 1--8.
  IEEE, Oxford (2019)

\bibitem{RedHat23}
Roguski, P., Prpic, M.: {V}ulnerability {E}xploitability e{X}change {(VEX)}
  {B}eta {F}iles {N}ow {A}vailable (2023), {R}ed {H}at {B}log, available online
  in August 2025:
  \url{https://www.redhat.com/en/blog/vulnerability-exploitability-exchange-vex-beta-files-now-available}

\bibitem{Ruohonen25ICTSSa}
Ruohonen, J.: {A} {T}ime {S}eries {A}nalysis of {A}ssertions in the {L}inux
  {K}ernel. In: Proceedings of the 37th International Conference on Testing
  Software and Systems (ICTSS 2025). pp. 3--15. Springer, Limassol (2025)

\bibitem{Ruohonen25DSVEX}
Ruohonen, J., Abdullahi, S., Tiwari, A.: {A} {R}eplication {P}ackage for a
  {P}aper {E}ntitled ``{V}ulnerability {P}atching {A}cross {S}oftware
  {P}roducts and {S}oftware {C}omponents: {A} {C}ase {S}tudy of {R}ed {H}at's
  {P}roduct {P}ortfolio'' (2025), {Z}enodo, available online:
  \url{https://doi.org/10.5281/zenodo.16987683}

\bibitem{Ruohonen15IWPSE}
Ruohonen, J., Hyrynsalmi, S., Lepp\"anen, V.: {S}oftware {E}volution and {T}ime
  {S}eries {V}olatility: {A}n {E}mpirical {E}xploration. In: Proceedings of the
  14th International Workshop on Principles of Software Evolution (IWPSE 2015).
  pp. 56--65. ACM, Bergamo (2015)

\bibitem{Ruohonen15COSE}
Ruohonen, J., Hyrynsalmi, S., Lepp\"anen, V.: {T}he {S}igmoidal {G}rowth of
  {O}perating {S}ystem {S}ecurity {V}ulnerabilities: {A}n {E}mpirical
  {R}evisit. Computers \& Security  55,  1--20 (2015)

\bibitem{Ruohonen16GIQ}
Ruohonen, J., Hyrynsalmi, S., Lepp\"anen, V.: {A}n {O}utlook on the
  {I}nstitutional {E}volution of the {E}uropean {U}nion {C}yber {S}ecurity
  {A}pparatus. Government Information Quarterly  33(4),  746--756 (2016)

\bibitem{Sion24}
Sion, L., {Van Landuyt}, D., Yskout, K., Verreydt, S., , Joosen, W.: {CTAM}:
  {A} {T}ool for {C}ontinuous {T}hreat {A}nalysis and {M}anagement. In:
  Sadovykh, A., Truscan, D., Mallouli, W., Cavalli, A.R., Seceleanu, C.,
  Bagnato, A. (eds.) {C}yber{S}ecurity in a {D}ev{O}ps {E}nvironment: {F}rom
  {R}equirements to {M}onitoring, pp. 195--223. Springer, Cham (2024)

\bibitem{Wochnik25}
Wochnik, J., Gr\"aupner, O.S., Spranger, M., Hummert, C.: {R}egarding the
  {E}xponential {G}rowth of {S}ecurity {V}ulnerabilities. In: Proceedings of
  the 23rd International Conference on Security and Management and Wireless
  Networks (CSCE 2024). pp. 329--343. Springer, Las Vegas (2025)

\bibitem{Xu22}
Xu, C., Chen, B., Lu, C., Huang, K., Peng, X., Liu, Y.: {T}racking {P}atches
  for {O}pen {S}ource {S}oftware {V}ulnerabilities. In: Proceedings of the 30th
  ACM Joint European Software Engineering Conference and Symposium on the
  Foundations of Software Engineering (ESEC/FSE 2022). pp. 860--871. ACM,
  Singapore (2022)

\end{thebibliography}

\end{document}